\documentclass[11pt]{article}

\usepackage[final]{acl}
\usepackage{times}
\usepackage{latexsym}

\usepackage[T1]{fontenc}
\usepackage[utf8]{inputenc}

\usepackage{microtype}

\usepackage{inconsolata}

\usepackage{graphicx}
\usepackage{amsmath}
\usepackage{amsfonts}
\usepackage{array}

\usepackage{booktabs}
\usepackage{multirow}
\usepackage{tabularx}

\usepackage{listings}
\usepackage{xcolor}
\usepackage[table]{xcolor}
\usepackage{makecell}

\usepackage[most]{tcolorbox}
\usepackage{pifont}
\newtcolorbox{gptprompt}{
  colback=gray!5,
  colframe=gray!50,
  colbacktitle=gray!50,
  coltitle=white,
  boxrule=0.4pt,
  arc=2pt,
  left=4pt,right=4pt,top=4pt,bottom=4pt,
  fontupper=\footnotesize,
  fonttitle=\bfseries,
  before upper={%
    \setlength{\parindent}{0pt}%
    \setlength{\parskip}{4pt}%
    \raggedright
    \obeylines
  },
  breakable,
  title=Prompt,
}
\usepackage{fvextra}
\usepackage[most]{tcolorbox}
\usepackage{xcolor}
\usepackage{listings}
\usepackage{fontawesome5}

\lstdefinestyle{promptbox}{
    basicstyle=\ttfamily\scriptsize,
    breaklines=true,
    breakatwhitespace=false,
    columns=fullflexible,
    keepspaces=true,
    showstringspaces=false,
    frame=none
}

\title{
PhysWave: Physics-Guided Latent Diffusion Models for Controllable Spatial Audio Generation
}

\author{
  \textbf{Lingfeng Yao\textsuperscript{1}},
  \textbf{Chenpei Huang\textsuperscript{1}},
  \textbf{Xingke Yang\textsuperscript{1}},
  \textbf{Ziye Geng\textsuperscript{1}},
\\
  \textbf{Changqing Luo\textsuperscript{1}},
  \textbf{Hao Wang\textsuperscript{2}},
  \textbf{Jiang Liu\textsuperscript{3}},
  \textbf{Miao Pan\textsuperscript{1}}
\\
\\
  \textsuperscript{1}University of Houston
  \textsuperscript{2}Stevens Institute of Technology
  \textsuperscript{3}Waseda University
\\
    \texttt{
      \{lyao12, mpan2\}@uh.edu
    }
\\
    \small{
      \href{https://lingfengyao.github.io/PhysWave/}
      {\faGlobe\ \textbf{Project Page}}
    }
}

\begin{document}
\maketitle
\begin{abstract}
Text-to-spatial audio generation, such as text-to-First-Order Ambisonics (FOA), provides a convenient way to create spatial audio for billion-dollar gaming and film industries.
However, existing text-to-FOA methods are largely data-driven and may produce audio that violates acoustic relations between source direction and distance. They also separate descriptive and parametric control, forcing users to trade usability for precision. In this paper, we present \textit{PhysWave}, a physics-guided latent diffusion model for controllable text-to-FOA generation. PhysWave unifies natural-language and trajectory control through a shared waypoint-caption representation, and augments diffusion training with two differentiable acoustic priors: spherical-harmonic direction consistency and inverse-square distance consistency. To support dynamic spatial generation, we further construct a 300K-clip FOA dataset with diverse sound categories and source trajectories. Extensive results show that the proposed priors help PhysWave generate spatially consistent FOA audio while maintaining competitive audio quality. Further analyses show that these physics priors improve spatial consistency during training and can also be used as inference-time guidance for training-free spatial refinement.
\end{abstract}

\section{Introduction}
Spatial audio generation is an essential part of the billion-dollar gaming and film industries, especially for immersive virtual reality (VR), augmented reality (AR) gaming, and 3D/4D movies, which require audio to convey a three-dimensional sound field. 
First-Order Ambisonics (FOA)~\citep{zotter2019ambisonics, gerzon1985ambisonics, malham19953} provides a practical format for this purpose. Unlike channel-based formats tied to fixed playback layout, FOA represents the sound field with four spherical-harmonic channels and can be decoded to headphones, loudspeaker arrays, and other playback systems. This device-agnostic property makes FOA a natural format for spatial audio generation. However, creating FOA audio remains challenging, which typically requires specialized tools, manual scene design, and spatial audio expertise. Those requirements limit FOA's use by non-expert users. That motivates text-conditioned FOA generation, where users can create spatial audio from natural-language descriptions.
\iffalse
Immersive applications such as virtual reality (VR), augmented reality (AR), games, and cinematic media require audio that conveys a three-dimensional sound field. First-Order Ambisonics (FOA)~\citep{zotter2019ambisonics, gerzon1985ambisonics, malham19953} provides a practical format for this purpose. Unlike channel-based formats tied to fixed playback layout, FOA represents the sound field with four spherical-harmonic channels and can be decoded to headphones, loudspeaker arrays, and other playback systems. This device-agnostic property makes FOA a natural format for spatial audio generation. However, creating FOA audio usually requires specialized tools, manual scene design, and spatial audio expertise, which limits its use by non-expert users. This motivates text-conditioned FOA generation, where users can create spatial audio from natural-language descriptions.
\fi

Text-to-audio generation has made strong progress in producing audio that faithfully matches text prompts. For example, monaural text-to-audio methods~\citep{liu2023audioldm, liu2024audioldm, huang2023make, majumder2024tango} mainly focused on audio quality and semantic alignment, but they did not model spatial structure. Stereo audio generation~\citep{sun2024both, feng2025audiospa, zhao2026dualspec} introduced spatial cues such as interaural time difference (ITD) and interaural level difference (ILD), providing a sense of left-right localization. However, two-channel audio is limited in representing a full three-dimensional sound field. Recent work has therefore moved toward FOA generation. ImmerseDiffusion~\citep{heydari2025immersediffusion} studied text-conditioned FOA synthesis for static sound scenes, and SonicMotion~\citep{templin2025generating} extended FOA generation to moving sound sources. Despite this progress, current text-to-FOA methods still have two major limitations.
\iffalse
Text-to-audio generation has made strong progress in generating audio that matches text prompts. Monaural text-to-audio methods~\citep{liu2023audioldm, liu2024audioldm, huang2023make, majumder2024tango} mainly focus on audio quality and semantic alignment, but they do not model spatial structure. Stereo audio generation~\citep{sun2024both, feng2025audiospa, zhao2026dualspec} introduces spatial cues such as interaural time difference (ITD) and interaural level difference (ILD), providing a sense of left-right localization. However, two-channel audio is limited in representing a full three-dimensional sound field. Recent work has therefore moved toward FOA generation. ImmerseDiffusion~\citep{heydari2025immersediffusion} studies text-conditioned FOA synthesis for static sound scenes, and SonicMotion~\citep{templin2025generating} extends FOA generation to moving sound sources. Despite this progress, current text-to-FOA methods still face two main limitations.
\fi

\paragraph{Limitation 1: Physical consistency cannot be guaranteed.}
FOA audio follows known acoustic relations. For a source arriving from azimuth $\theta$ and elevation $\phi$, the FOA directional channels should match the first-order spherical-harmonic encoding. For a source at distance $r$, the received energy should vary with distance, which is commonly modeled by an inverse-square relation in a free field. Existing FOA generation methods mainly learn these relations implicitly from data. As a result, a generated clip may sound plausible but still encode an incorrect source direction in its FOA channels, or produce an energy envelope that does not match the source-listener distance. Since these relations are known and differentiable, they can be leveraged as explicit physical priors during training and sampling.

\iffalse
\paragraph{Limitation 1: Physical consistency is only learned implicitly.}
FOA audio follows known acoustic relations. For a source arriving from azimuth $\theta$ and elevation $\phi$, the FOA directional channels should match the first-order spherical-harmonic encoding. For a source at distance $r$, the received energy should vary with distance, which is commonly modeled by an inverse-square relation in a free field. Existing FOA generation methods mainly learn these relations implicitly from data. As a result, a generated clip may sound plausible but still encode an incorrect source direction in its FOA channels, or produce an en energy envelope that does not follow the source-listener distance. Since these relations are known and differentiable, they can be used as explicit physical priors during training and sampling.
\fi

\paragraph{Limitation 2: User interfaces are separated.}
Current FOA generation methods often provide two separate control interfaces~\citep{heydari2025immersediffusion, templin2025generating}. A \emph{parametric} mode accepts numerical spatial parameters such as azimuth, elevation, distance, and motion, enabling precise control but requiring users to specify low-level spatial variables. A \emph{descriptive} mode accepts natural-language prompts, which are easy to use but usually provide coarse control over direction, distance, and time-varying motion. This separation forces users to choose between ease of use and fine-grained spatial control. A unified interface is therefore needed to support both interaction modes within the same generation model.

To address these limitations, we propose \textit{PhysWave}, a physics-guided latent diffusion model for text-to-FOA generation. PhysWave augments the standard diffusion denoising objective with two differentiable acoustic losses: (i) a \emph{spherical-harmonic direction consistency} loss that aligns FOA cross-channel relations with the target source direction, and (ii) an \emph{inverse-square distance consistency} loss that regularizes the generated energy envelope according to the source-listener distance. These losses provide explicit physical priors and guide the model to generate FOA audio that matches the requested spatial trajectory.
PhysWave also introduces a unified \emph{waypoint-caption} representation for spatial control. A user can describe a sound scene in natural language, which is parsed into a non-spatial acoustic caption and an editable waypoint trajectory, or directly provide waypoints for fine-grained control. Both input forms are processed by the same diffusion transformer (DiT)-based model~\citep{peebles2023scalable}. To support training and evaluation, we construct a 300K-clip dynamic FOA dataset with diverse sound categories and source trajectories. Experiments show that PhysWave improves source direction and distance consistency while maintaining competitive audio quality. Further analyses demonstrate that the proposed physics priors can also be applied as inference-time guidance for training-free spatial refinement. Our salient contributions are summarized as follows.

\begin{itemize}
    \item We propose \textbf{PhysWave}, a novel physics-guided latent diffusion framework for controllable text-to-FOA generation.
    \item We introduce \textbf{differentiable acoustic priors} for spherical-harmonic direction consistency and inverse-square distance consistency, which improve spatial consistency during training and enable inference-time spatial refinement without retraining.
    \item We design a \textbf{waypoint-caption representation} that unifies natural-language descriptions and editable spatial trajectories within the same generation model.
    \item We construct a \textbf{300K-clip dynamic FOA dataset} with diverse sound categories and source trajectories, supporting both large-scale training and fine-grained evaluations.
\end{itemize}

\section{Related Work}
\subsection{Text-to-Audio Generation}
Existing audio generation methods can be grouped by output format: monaural, stereo/binaural, and FOA spatial audio. Monaural text-to-audio methods, such as AudioLDM series~\citep{liu2023audioldm, liu2024audioldm}, Make-An-Audio series~\citep{huang2023make, huang2023make2}, TANGO series~\citep{ghosal2023tango, majumder2024tango}, and Stable Audio~\citep{evans2024fast}, have improved audio fidelity and semantic alignment. However, their outputs do not encode source direction, distance, or motion. Stereo audio generation methods introduce spatial cues through two-channel audio. AudioSpa~\citep{feng2025audiospa} and DualSpec~\citep{zhao2026dualspec} study text-guided binaural generation, while SpatialSonic~\citep{sun2024both} supports controllable stereo audio generation from language and other modalities. These methods improve spatial perception, but binaural audio remains a listener-centric format and does not fully represent the three-dimensional sound field.

FOA has recently been used as a device-agnostic format for spatial audio generation. Diff-SAGe~\citep{kushwaha2025diff} studies FOA generation from sound category and source location. ImmerseDiffusion~\citep{heydari2025immersediffusion} introduces text-conditioned FOA generation for static sound scenes, and SonicMotion~\citep{templin2025generating} extends FOA generation to moving sound sources. These works show the potential of FOA for immersive audio generation. However, existing FOA generators remain mainly data-driven. They learn spatial structure from paired training data without explicitly using acoustic relations in the training objective. They also often separate descriptive and parametric control, which limits either ease of use or precise spatial control. In contrast, PhysWave introduces differentiable acoustic priors for FOA generation and uses a unified waypoint-caption representation for both natural-language and trajectory-based control. A detailed comparison is provided in Appendix~\ref{app:comparison}.

\subsection{Physics-Guided Generative Models}
Physics-guided learning incorporates known physical relations into neural network training. Physics-Informed Neural Networks (PINNs)~\citep{raissi2019physics} penalize violations of governing equations, and related ideas have been extended to generative models. For example, physics-informed diffusion models~\citep{bastek2025physicsinformed} use residual losses or physical constraints to improve generation in scientific domains, including flow fields~\citep{shu2023physics}, temperature downscaling~\citep{rosu2025pde}, and infrared imagery~\citep{mao2026pid}. In acoustics, physics-informed learning has been studied for sound field estimation and reconstruction, often using priors based on the wave equation, Helmholtz equation, or physics-constrained kernels~\citep{olivieri2024physics, ribeiro2024sound, koyama2025physics}. These works focus on inverse reconstruction from measurement rather than text-conditioned audio generation. PhysWave differs by applying simple differentiable acoustic priors to controllable FOA generation, where spherical-harmonic direction consistency and inverse-square distance consistency directly match the spatial cues specified by source trajectories.

\section{Spatial Audio Dataset and Representation}
\label{sec:dataset_construction}
\begin{figure}[t]
    \centering
    \includegraphics[width=0.49\textwidth]{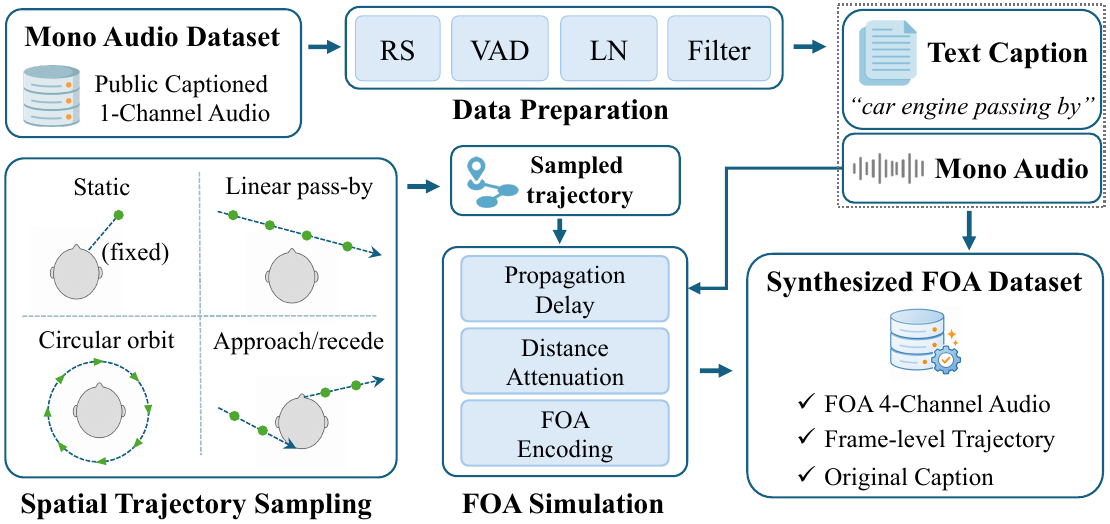}
    \caption{Dataset construction pipeline. Captioned monaural audio is paired with sampled trajectories and rendered into FOA audio through simulation.}
    \label{fig:dataset_pipeline}
\end{figure}

Controllable FOA generation requires paired data that specify both the acoustic event and its listener-relative motion. However, real FOA recordings with accurate source trajectories are difficult to collect at scale. We therefore construct a large synthetic dataset by spatializing captioned monaural audio along sampled source trajectories. As shown in Figure~\ref{fig:dataset_pipeline}, our pipeline consists of three stages: monaural audio preparation, spatial trajectory sampling, and physics-based FOA simulation.

\paragraph{Data preparation.}
We collect monaural clips from AudioCaps~\citep{audiocaps}, WavCaps~\citep{mei2023wavcaps}, and Clotho~\citep{drossos2020clotho}. Each clip is processed in four steps. First, we apply resampling (RS) to convert all audio to $16$ kHz. Second, we use voice activity detection (VAD) to select $10$-second segments with sufficient active content. Third, we perform loudness normalization (LN) to reduce scale variation across sources.  Finally, we filter weakly matched caption-audio pairs using CLAP similarity. More details are provided in Appendix~\ref{app:dataset}.

\begin{figure*}[t]
    \centering
    \includegraphics[width=\textwidth]{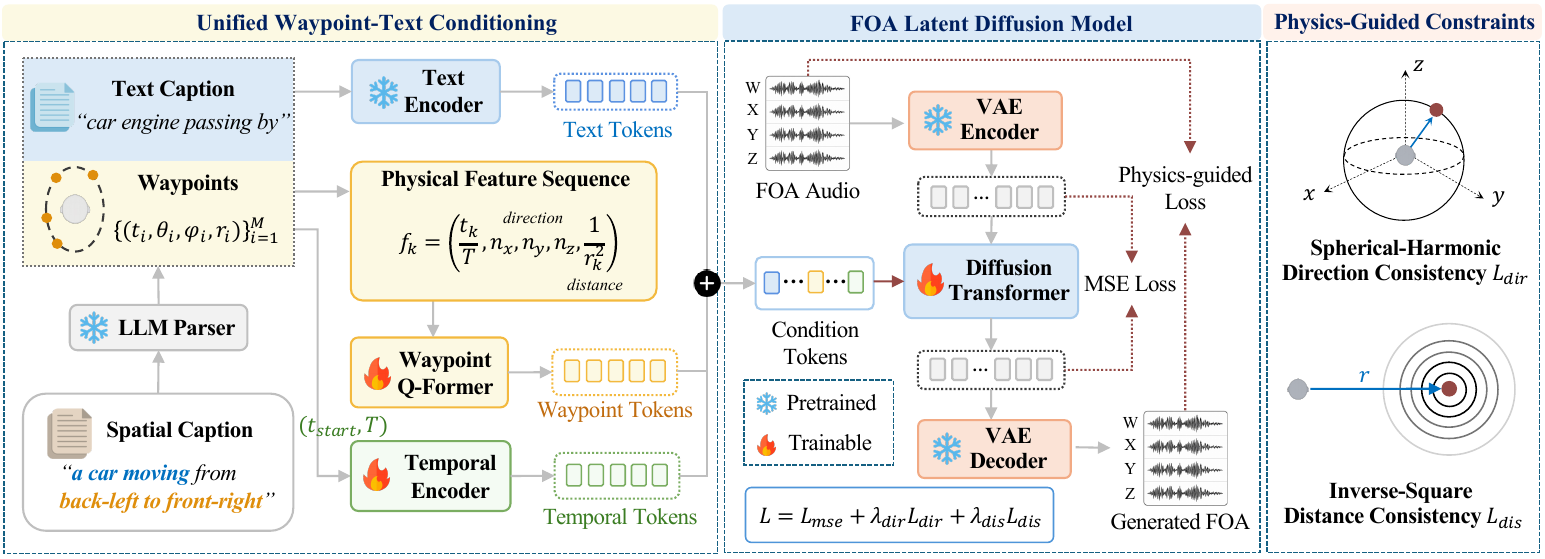}
    \caption{Overview of PhysWave. A spatial caption is parsed into an acoustic caption and a waypoint trajectory. Text, waypoint, and temporal tokens jointly condition an FOA latent diffusion model, which is trained with latent denoising loss and physics-consistency losses on the decoded FOA waveform.}
    \label{fig:framework}
\end{figure*}

% Define FOA format and 
\paragraph{FOA representation.}
FOA represents a three-dimensional sound field using four spherical-harmonic channels. We denote an FOA waveform as
\begin{equation}
    \mathbf{a}(t) = [W(t), X(t), Y(t), Z(t)]^\top ,
\end{equation}
where $W$ is the omnidirectional component, and $(X,Y,Z)$ encode directional projections along the front--back, left--right, and up--down axes. For a point source with pressure signal $s(t)$ arriving from azimuth $\theta$ and elevation $\phi$, the FOA encoding is
\begin{equation}
\begin{aligned}
    W(t) &= \frac{1}{\sqrt{2}}s(t), &
    X(t) &= s(t)\cos\phi\cos\theta, \\
    Y(t) &= s(t)\cos\phi\sin\theta, &
    Z(t) &= s(t)\sin\phi .
\end{aligned}
\label{eq:foa_encoding}
\end{equation}
The four channels are therefore not independent audio streams. Their cross-channel relations encode the spatial direction, which is essential for trajectory-controllable FOA generation.

\paragraph{Trajectory representation.}
We describe source motion by a listener-relative trajectory
$(\theta(t), \phi(t), r(t))$, where $\theta(t)$ is the azimuth, $\phi(t)$ is the elevation, and $r(t)$ is the source-listener distance. To make trajectories compact and editable, each trajectory is specified by a waypoint sequence
\begin{equation}
    \mathbf{W}=\{(t_i,\theta_i,\phi_i,r_i)\}_{i=1}^{M},
\end{equation}
where $M$ controls the temporal granularity. During simulation, we linearly interpolate the waypoint sequence to obtain sample-level directions and distances. 
As illustrated in Figure~\ref{fig:dataset_pipeline}, we sample trajectories from four motion families: \emph{static} sources with fixed direction and distance, \emph{circular orbits} with changing direction and nearly constant distance, \emph{approach/recede} trajectories with strong radial distance changes, and \emph{linear pass-bys} with coupled direction and distance changes.

\paragraph{FOA simulation.}
Given a monaural clip $s(t)$ and a sampled trajectory $(\theta(t), \phi(t), r(t))$, we render FOA audio with an acoustic simulation pipeline. \emph{(i) Propagation delay} resamples the source signal on the retarded-time grid $t_{\mathrm{emit}}=t-r(t)/c$, where $c$ is the speed of sound. This makes each output sample depend on the pressure emitted at the corresponding source distance and induces Doppler-like frequency changes under radial motion. \emph{(ii) Distance attenuation} scales the delayed signal by $1/r(t)$ to model free-field pressure decay. \emph{(iii) FOA encoding} projects the resulting pressure signal onto the four spherical-harmonic channels using Eq.~\eqref{eq:foa_encoding}. The final synthesized dataset contains triples of \{FOA audio, frame-level trajectory, original caption\}.

\section{Physics-Guided Latent Diffusion Model}
\subsection{Overview}
Our goal is to generate FOA audio that is both semantically aligned with a text prompt and spatially consistent with a user-specified source trajectory. As shown in Figure~\ref{fig:framework}, PhysWave consists of three main components: unified waypoint-caption conditioning, an FOA latent diffusion backbone, and physics-guided training objectives.

Given a spatial caption, a large language model (LLM) parser decomposes it into two structured conditions: a text caption that describes the sound event and a waypoint sequence specifying the source motion. The text caption is encoded into text tokens. The waypoint sequence is converted into physical trajectory features and encoded into waypoint tokens. A temporal encoder further maps temporal information into temporal tokens. These token sets are concatenated as the joint condition for the diffusion model.

PhysWave follows a latent diffusion design. A four-channel FOA waveform $\mathbf{a}$ is compressed into a continuous latent $\mathbf{x}$ by an FOA variational autoencoder (VAE) encoder. A DiT then denoises the latent under the joint condition. Unlike standard latent diffusion training, which applies supervision only in latent space, we decode the predicted clean latent back to an FOA waveform during training and apply differentiable physical constraints to the decoded signal. This encourages the model to match both the acoustic caption and the target source trajectory.

\subsection{Spatial Physical Priors}
\label{sec:spatial_physical_priors}
A source trajectory provides two physical cues that a faithful FOA signal should follow: the instantaneous direction of arrival and the distance-induced energy envelope. We formalize them as two physical priors, which are used for both waypoint conditioning in Section~\ref{sec:unified_conditioning} and physics-guided training in Section~\ref{sec:physics_objective}.

\paragraph{Spherical-harmonic direction prior.}
For a source arriving from azimuth $\theta(t)$ and elevation $\phi(t)$, the listener-relative unit direction is
\begin{equation}
    \mathbf{n}(t)
    =
    \begin{bmatrix}
        n_x(t) \\
        n_y(t) \\
        n_z(t)
    \end{bmatrix}
    =
    \begin{bmatrix}
        \cos\phi(t)\cos\theta(t) \\
        \cos\phi(t)\sin\theta(t) \\
        \sin\phi(t)
    \end{bmatrix}.
    \label{eq:direction_vector}
\end{equation}
We use $\mathbf{n}(t)$ as the per-frame direction target. As defined in Eq.~\eqref{eq:foa_encoding}, FOA represents direction through cross-channel relations between the omnidirectional channel $W$ and the directional channels $(X,Y,Z)$. A spatially consistent FOA signal should therefore recover the target direction from this cross-channel structure.

\paragraph{Inverse-square distance prior.}
Free-field propagation attenuates pressure approximately by $1/r(t)$, where $r(t)$ is the source-listener distance. The corresponding distance-induced energy component follows an inverse-square profile $1/r(t)^2$. Since source pressure $s(t)$ scales the FOA channels in Eq.~\eqref{eq:foa_encoding}, this profile provides a trajectory-conditioned attenuation prior for generated FOA audio. We use it to constrain the relative temporal energy trend of the generated signal while leaving the absolute source loudness unconstrained.

\subsection{Unified Waypoint-Caption Conditioning}
\label{sec:unified_conditioning}
A spatial caption is first parsed by the frozen LLM parser into a text caption and a waypoint sequence, as illustrated in Figure~\ref{fig:framework}. The text caption specifies the sound event and is encoded by a pretrained T5 encoder~\citep{raffel2020exploring} into \textit{text tokens}. The waypoint sequence specifies the source motion:
\begin{equation}
    \mathbf{W}=\{(t_i,\theta_i,\phi_i,r_i)\}_{i=1}^{M}.
\end{equation}
To expose the physical structure of this trajectory to the model, we interpolate the waypoints into a dense per-frame feature sequence:
\begin{equation}
    \mathbf{f}_k
    =
    \big[
        \underbrace{\tfrac{t_k}{T}}_{\text{time}},\;
        \underbrace{n_x(t_k),\, n_y(t_k),\, n_z(t_k)}_{\text{direction}},\;
        \underbrace{\tfrac{1}{r(t_k)^2}}_{\text{distance}}
    \big],
    \label{eq:traj_feature}
\end{equation}
where $T$ is the clip duration, $(n_x,n_y,n_z)$ is the unit direction vector from Eq.~\eqref{eq:direction_vector}, and $1/r(t_k)^2$ is the inverse-square distance profile.

A waypoint Q-Former attends a small set of learnable queries to $\{\mathbf{f}_k\}$ and produces \textit{waypoint tokens}. In parallel, a temporal encoder embeds the clip start time and total duration $(t_{\mathrm{start}}, T)$ into \textit{temporal tokens}. We concatenate the text tokens, waypoint tokens, and temporal tokens into the joint condition $\mathbf{c}$, which drives the DiT through cross-attention.

\subsection{FOA Latent Diffusion Backbone}
We use a four-channel FOA VAE to compress FOA waveforms into continuous latents. Because FOA channels jointly encode spatial structure, we pretrain the VAE with standard waveform reconstruction losses and the spherical-harmonic direction consistency loss $\mathcal{L}_{\mathrm{dir}}$ in Eq.~\eqref{eq:loss_dir}, applied to the VAE output. VAE training details are provided in Appendix~\ref{app:foa_vae}. After pretraining, VAE is frozen during diffusion training.

Given a clean latent $\mathbf{x}$, we sample Gaussian noise $\boldsymbol{\epsilon}\sim\mathcal{N}(0,\mathbf{I})$ and construct the noisy latent
\begin{equation}
    \mathbf{x}_t = \alpha_t \mathbf{x} + \sigma_t \boldsymbol{\epsilon},
\end{equation}
where $\alpha_t$ and $\sigma_t$ follow a cosine noise schedule. The DiT predicts the velocity target
\begin{equation}
    \mathbf{v}_t = \alpha_t \boldsymbol{\epsilon} - \sigma_t \mathbf{x},
\end{equation}
conditioned on $\mathbf{c}$. The standard latent diffusion objective is
\begin{equation}
    \mathcal{L}_{\mathrm{mse}}
    =
    \mathbb{E}_{\mathbf{x},\boldsymbol{\epsilon},t}
    \!\left[
    \big\|
    \hat{\mathbf{v}}_\theta(\mathbf{x}_t,t,\mathbf{c})
    -
    \mathbf{v}_t
    \big\|_2^2
    \right].
    \label{eq:diffusion_loss}
\end{equation}

\subsection{Physics-Guided Training Objective}
\label{sec:physics_objective}
To improve the physical consistency of generated FOA audio, we compute auxiliary losses on the decoded waveform rather than on the latent. These losses encourage the generated waveform to follow the direction and distance relations specified by the input trajectory. Given the velocity prediction $\hat{\mathbf{v}}_\theta$, we recover the predicted clean latent $\hat{\mathbf{x}}_0 = \alpha_t \mathbf{x}_t - \sigma_t \hat{\mathbf{v}}_\theta(\mathbf{x}_t,t,\mathbf{c})$ and decode it with the frozen VAE decoder $\mathcal{D}$ to obtain the predicted FOA waveform $\hat{\mathbf{a}} = \mathcal{D}(\hat{\mathbf{x}}_0)$. We then apply one consistency loss for each physical prior.

\paragraph{Spherical-harmonic direction consistency.}
For each frame $k$, we estimate a short-time FOA intensity vector~\citep{adavanne2018direction} from the generated waveform:
\begin{equation}
    \hat{\mathbf{I}}_k =
    \big[
    \langle \hat{W}_k \hat{X}_k\rangle,\;
    \langle \hat{W}_k \hat{Y}_k\rangle,\;
    \langle \hat{W}_k \hat{Z}_k\rangle
    \big]^\top,
\end{equation}
where $\langle\cdot\rangle$ denotes averaging within frame $k$. The direction loss aligns this vector with the target direction $\mathbf{n}_k$:
\begin{equation}
    \mathcal{L}_{\mathrm{dir}}
    =
    \frac{1}{K}\sum_{k=1}^{K}
    \left(
    1-
    \frac{\hat{\mathbf{I}}_k^\top \mathbf{n}_k}
    {\|\hat{\mathbf{I}}_k\|\,\|\mathbf{n}_k\|+\varepsilon}
    \right).
    \label{eq:loss_dir}
\end{equation}
where $K$ is the number of frames and $\varepsilon$ is a small constant for numerical stability.

\paragraph{Inverse-square distance consistency.}
We measure the distance-induced energy envelope using the omnidirectional FOA channel. For each frame $k$, we compute the generated $W$-channel energy:
\begin{equation}
    \hat{E}_k
    =
    \langle \hat{W}_k^2\rangle .
\end{equation}
The target energy profile is $E^{*}_k=1/r_k^2$. Since absolute loudness depends on the source content, we compare only the normalized log-energy shape. Let $\widetilde{u}_k = \log(u_k+\varepsilon) - \tfrac{1}{K}\sum_j \log(u_j+\varepsilon)$. The distance consistency loss is
\begin{equation}
    \mathcal{L}_{\mathrm{dist}}
    =
    \frac{1}{K}\sum_{k=1}^{K}
    \big(
    \widetilde{\hat{E}}_k
    -
    \widetilde{E}^{*}_k
    \big)^2.
    \label{eq:loss_dist}
\end{equation}

\paragraph{Final objective.}
The final objective combines latent denoising with the two physics-guided constraints:
\begin{equation}
    \mathcal{L}
    =
    \mathcal{L}_{\mathrm{mse}}
    +
    \lambda_{\mathrm{dir}}\mathcal{L}_{\mathrm{dir}}
    +
    \lambda_{\mathrm{dist}}\mathcal{L}_{\mathrm{dist}},
    \label{eq:total_loss}
\end{equation}
where $\lambda_{\mathrm{dir}}$ and $\lambda_{\mathrm{dist}}$ control the strength of the direction and distance constraints.

\section{Experiments}
\label{sec:experiments}
\subsection{Implementation Details}
\paragraph{Dataset.}
We use the synthetic FOA dataset described in Section~\ref{sec:dataset_construction}. Each monaural clip is paired with a sampled listener-relative source trajectory and rendered into four-channel FOA audio. Appendix~\ref{app:dataset} provides further details.

\paragraph{Model.}
We use Qwen3.5-4B~\citep{qwen3.5} as the LLM parser to convert spatial captions into structured waypoint-caption conditions. Our generative backbone follows Stable Audio Open~\citep{evans2025stable}: we adapt its continuous DAC-based VAE from stereo to four-channel FOA WXYZ input/output, and use its 24-layer DiT denoiser with hidden size 768 and 12 attention heads. Text captions are encoded by a pretrained T5-base encoder~\citep{raffel2020exploring} with a maximum length of 128 tokens. Waypoints are encoded by a lightweight Q-Former-style encoder into 16 learnable query tokens, and temporal scalars are encoded with scalar embedding layers. Further details are provided in Appendix~\ref{app:model}.

\subsection{Evaluation Metrics}
We evaluate generated FOA audio from two aspects: audio quality and spatial fidelity.

\paragraph{Audio quality.}
Following standard text-to-audio evaluation protocols, we report CLAP score, Fréchet Distance (FD), CLAP-based Fréchet Audio Distance (FAD$_{\mathrm{CLAP}}$), Inception Score (IS), and Kullback--Leibler divergence (KL). These metrics measure semantic alignment and perceptual quality of generated audio. Since they are mainly designed for single-channel audio, we compute them on the omnidirectional $W$ channel of the FOA signal.

\begin{table*}[t]
\centering
\small
\setlength{\tabcolsep}{4pt}
\renewcommand{\arraystretch}{1.05}
\resizebox{\textwidth}{!}{%
\begin{tabular}{l|ccccc|cccc}
\toprule
\multirow{2}{*}{\textbf{Variant}}
& \multicolumn{5}{c|}{\textbf{Audio quality}}
& \multicolumn{4}{c}{\textbf{Spatial fidelity}} \\
\cmidrule(lr){2-6}\cmidrule(lr){7-10}
& FD $\downarrow$
& FAD$_{\mathrm{CLAP}}$ $\downarrow$
& CLAP $\uparrow$
& KL $\downarrow$
& IS $\uparrow$
& Static ($^\circ$) $\downarrow$
& Moving ($^\circ$) $\downarrow$
& InvSq Err. (dB) $\downarrow$
& InvSq Corr. $\uparrow$ \\
\midrule
Trajectory text
& 23.13 & 0.20 & 0.32 & 1.74 & 8.05
& 7.33 & 17.62 & 4.59 & 0.62 \\
PhysWave (w/o physics)
& 21.92 & 0.20 & 0.33 & 1.69 & 8.22
& 2.05 & 6.50 & 5.15 & 0.48 \\
PhysWave (w/o $\mathcal{L}_{\mathrm{dist}}$) 
& 21.79 & 0.20 & 0.33 & 1.68 & 8.15
& 1.75 & 4.81 & 4.79 & 0.59 \\
PhysWave (w/o $\mathcal{L}_{\mathrm{dir}}$)
& 21.34 & \textbf{0.20} & 0.33 & 1.67 & 8.25
& 2.47 & 6.60 & \textbf{3.78} & \textbf{0.74} \\
\rowcolor{blue!10}
\textbf{PhysWave}
& \textbf{21.22} & 0.21 & \textbf{0.33} & \textbf{1.66} & \textbf{8.31}
& \textbf{1.73} & \textbf{4.65} & 4.06 & 0.73 \\
\bottomrule
\end{tabular}}
\caption{Ablation study of PhysWave. All variants use the same generative backbone and caption condition. Angular errors are reported in degrees, and InvSq Err. is reported in dB.}
\label{tab:main_ablation}
\end{table*}

\paragraph{Spatial fidelity.}
We evaluate spatial quality in terms of direction consistency and distance consistency. For direction consistency, we estimate the direction of arrival (DoA) at each frame from the FOA intensity vector $(\langle WX\rangle,\langle WY\rangle,\langle WZ\rangle)$ and convert it to azimuth $\hat{\theta}$ and elevation $\hat{\phi}$. We then compute the spherical angular error~\citep{heydari2025immersediffusion} between the estimated and ground-truth directions using the haversine formula. 
For distance consistency, we evaluate whether the generated signal follows the distance attenuation profile specified by the ground-truth trajectory. Specifically, we compute the per-frame $W$-channel energy and compare it with the target inverse-square profile $1/r_k^2$. Since absolute loudness depends on the audio content, we convert both sequences to log scale and mean-center them before comparison. To reduce the effect of silent or low-energy frames, we apply an energy gate on the generated $W$-channel and compute the metrics over active frames. We report RMS error (InvSqErr, in dB), which measures profile mismatch, and Pearson correlation (InvSqCorr), which measures temporal agreement. Lower InvSqErr and higher InvSqCorr indicate better distance-consistent attenuation. Full metric definitions are provided in Appendix~\ref{app:metric}.

\subsection{Main Results}

\begin{table}[t]
\centering
\small
\setlength\tabcolsep{3pt}
\begin{tabular}{lccc}
\toprule
\textbf{Model}
& \textbf{Control}
& \textbf{Static} ($^\circ$) $\downarrow$
& \textbf{Moving} ($^\circ$) $\downarrow$ \\
\midrule
% ID-D & Static text     & 1.35 & --     \\
% ID-P & Static param.   & 1.12 & --     \\
% SM-D & Circular text   & --   & 29.22  \\
% SM-P & Circular param. & --   & 14.32  \\
ID-D & Static text     & 7.07 & --     \\
ID-P & Static param.   & 2.79 & --     \\
SM-D & Circular text   & 20.82   & 31.09  \\
SM-P & Circular param. & 2.41   & 19.22  \\
\midrule
\rowcolor{blue!10}
\textbf{PhysWave}  & Text + Waypoints & \textbf{1.73}    & \textbf{5.78} \\
\bottomrule
\end{tabular}
\caption{
Reference comparison with published results from prior FOA generation methods. ID and SM denote ImmerseDiffusion and SonicMotion, respectively. D and P denote descriptive and parametric conditioning.
}

\label{tab:spatial_fidelity}
\end{table}

\paragraph{Spatial fidelity.}
\label{sec:exp_spatial}
We compare PhysWave with two representative FOA generation methods: ImmerseDiffusion~\citep{heydari2025immersediffusion}, which focuses on static-source generation, and SonicMotion~\citep{templin2025generating}, which supports both static sources and circular source motion. Since their official implementations are not publicly available, we reproduce both methods and evaluate all models under the same evaluation protocol. Specifically, we compare ImmerseDiffusion and PhysWave on the same static-source test set, and compare SonicMotion and PhysWave on the same static-source and circular-motion test sets. As shown in Table~\ref{tab:spatial_fidelity}, PhysWave achieves the lowest DoA errors under both settings, with $1.73^\circ$ for static sources and $5.78^\circ$ for circular motion. In comparison, SonicMotion obtains circular-motion errors of $31.09^\circ$ with descriptive conditioning and $19.22^\circ$ with parametric conditioning, while ImmerseDiffusion obtains static-source errors of $7.07^\circ$ and $2.79^\circ$ under descriptive and parametric conditioning, respectively. These results suggest that PhysWave improves controllable moving-source localization while maintaining strong static-source accuracy.

\paragraph{Audio quality.}
\label{sec:exp_quality}
We evaluate semantic alignment and perceptual quality on the omnidirectional $W$ channel, following the FOA evaluation protocol of~\citet{templin2025generating}. Since standard text-to-audio metrics are mainly designed for monaural audio, the $W$ channel provides a natural way to evaluate the non-directional acoustic content of FOA audio. We compare PhysWave with state-of-the-art monaural text-to-audio models. Since these baselines do not generate FOA audio, we spatialize each generated monaural clip into FOA using the same trajectory-conditioned rendering pipeline described in Section~\ref{sec:dataset_construction}.
As shown in Table~\ref{tab:audio_quality}, PhysWave achieves competitive audio quality. It obtains the best KL score ($1.66$), while its FD ($21.22$), FAD$_{\mathrm{CLAP}}$ ($0.21$), CLAP score ($0.33$), and IS ($8.31$) remain comparable to strong monaural baselines. These results show that PhysWave adds controllable FOA spatialization while preserving the audio quality of the generated content.

\begin{figure*}[t]
    \centering
    \includegraphics[width=0.85\textwidth]{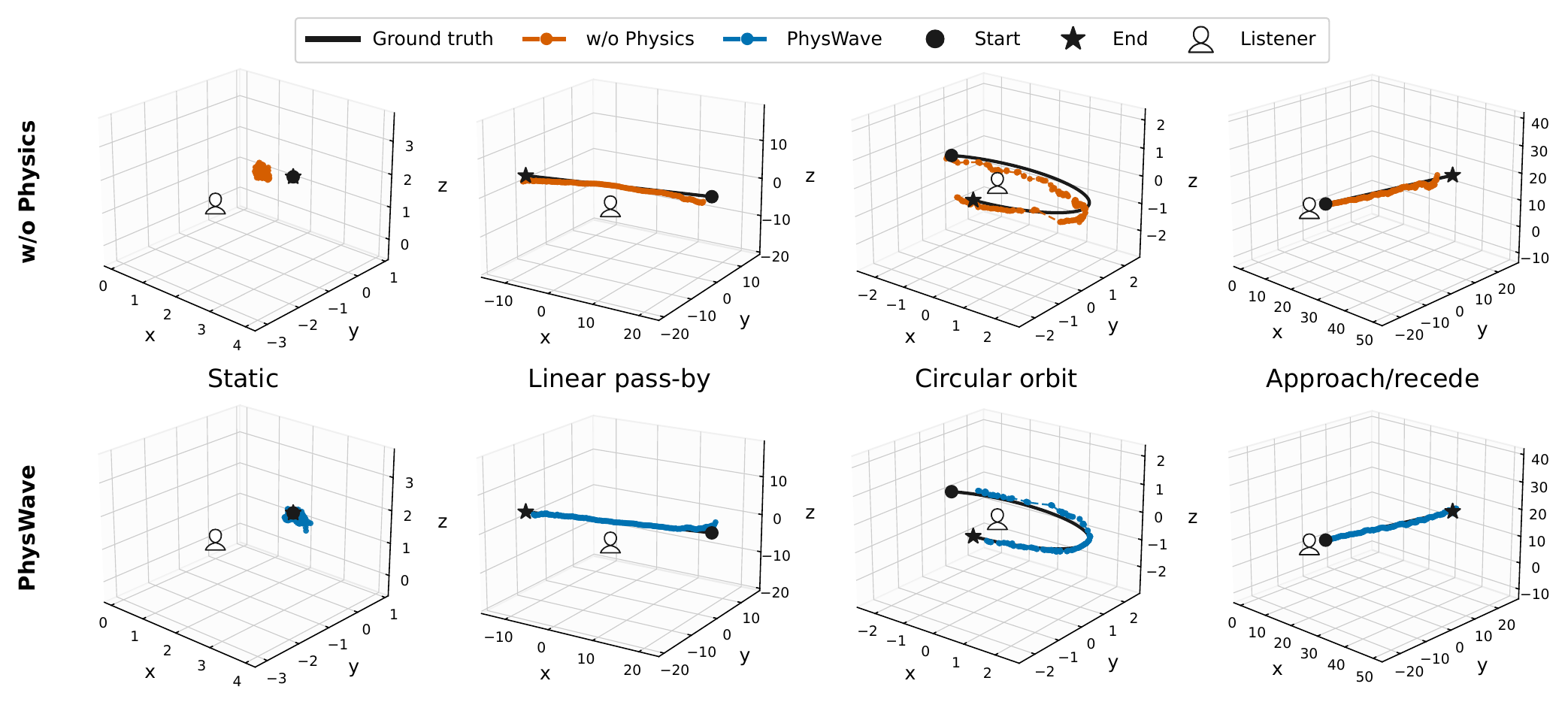}
    \caption{Qualitative trajectory comparison between PhysWave with and without physics-consistency losses.}
    \label{fig:qualitative_trajectory}
\end{figure*}

\begin{table}[t]
\centering
\small

\setlength\tabcolsep{1.7pt}
\begin{tabular}{l|ccccc}
\toprule
\textbf{Model}
& FD $\downarrow$
& FAD$_{\mathrm{CLAP}}$ $\downarrow$
& CLAP $\uparrow$
& KL $\downarrow$
& IS $\uparrow$ \\
\midrule
AudioLDM            & 31.67          & 0.33          & 0.33          & 2.37          & 6.96  \\
AudioLDM 2          & 24.13          & 0.12          & 0.33          & 2.11          & 8.62  \\
Make-An-Audio       & 15.44          & \textbf{0.11} & 0.37          & 1.93          & 8.79  \\
Make-An-Audio 2     & \textbf{13.86} & 0.14          & \textbf{0.40} & 1.73          & 10.85 \\
Stable Audio Open   & 30.84          & 0.31          & 0.30          & 2.30          & \textbf{11.09} \\
ImmerseDiffusion & 20.45 &0.14 &0.35 &1.66 &9.55 \\
SonicMotion         & 24.80          & 0.26          & 0.32          & 1.88          & 8.55  \\
\midrule
\rowcolor{blue!10}
\textbf{PhysWave }        & 21.22          & 0.21          & 0.33          & \textbf{1.66} & 8.31  \\
\bottomrule
\end{tabular}
\caption{Audio quality comparison on the omnidirectional $W$ channel.}
\label{tab:audio_quality}
\end{table}

\paragraph{Ablation study.}
We ablate the spatial conditioning interface and the physics consistency losses in Table~\ref{tab:main_ablation}. First, we compare waypoint conditioning with a text-based trajectory condition, where each trajectory is converted into a detailed natural-language description. Replacing waypoints with trajectory text sharply reduces spatial fidelity: static angular error increases from $2.05^\circ$ to $7.33^\circ$, and moving angular error increases from $6.50^\circ$ to $17.62^\circ$. This shows that free-form text is not precise enough to specify fine-grained source motion, including direction, distance, and temporal changes. In contrast, waypoints provide explicit geometric conditions, while the LLM parser bridges user text and structured waypoint inputs.
Second, the two physics losses improve different aspects of spatial fidelity. The direction loss reduces the moving angular error from $6.50^\circ$ to $4.81^\circ$ and the static angular error from $2.05^\circ$ to $1.75^\circ$, while having limited effects on the inverse-square metrics. The distance loss mainly improves distance consistency, increasing InvSq Corr. from $0.48$ to $0.74$ and reducing InvSq Err. from $5.15$ dB to $3.78$ dB. The full PhysWave model combines both losses, achieving the best static angular error ($1.73^\circ$), moving angular error ($4.65^\circ$) and strong distance consistency (InvSq Corr. $=0.73$). Audio quality remains stable across variants, indicating that the auxiliary physics losses improve spatial fidelity without degrading content quality.

\begin{table}[t]
\centering
\small
\setlength{\tabcolsep}{3pt}
\begin{tabular}{l|cc}
\toprule
\textbf{Parser}
& Spatial Faith. $\uparrow$
& Semantic Pres. $\uparrow$ \\
\midrule
Qwen3.5-0.8B
& 1.89 $\pm$ 0.20
& 3.40 $\pm$ 0.38 \\
Llama-3.2-3B-Instruct
& 2.40 $\pm$ 0.24
& 4.18 $\pm$ 0.29 \\
\rowcolor{blue!10}
\textbf{Qwen3.5-4B}
& \textbf{3.88} $\pm$ 0.28
& \textbf{4.72} $\pm$ 0.17 \\
\bottomrule
\end{tabular}
\caption{LLM parser evaluation on spatial captions. Scores are reported as mean $\pm$ standard deviation.}
\label{tab:parser_eval}
\end{table}

\begin{table}[t]
\centering
\small
\setlength{\tabcolsep}{3pt}
\begin{tabular}{lccc}
\toprule
\multirow{2}{*}{\textbf{Source}}
& MOS- & MOS- & MOS- \\
& Event $\uparrow$ & Trajectory $\uparrow$ & Realism $\uparrow$ \\
\midrule
Real FOA Recording & 4.89 & 4.56 & 4.79 \\
AudioLDM2 & 2.70 & 2.07 & 2.07 \\
Stable Audio Open & 3.40 & 2.21 & 2.51 \\
\rowcolor{blue!10}
\textbf{PhysWave} & \textbf{4.32} & \textbf{3.90} & \textbf{3.48} \\
\bottomrule
\end{tabular}
\caption{Subjective evaluation of generated FOA audio. MOS scores are obtained from 14 participants.}
\label{tab:subjective_eval}
\end{table}
\paragraph{Subjective evaluation.}
To assess the perceptual quality of generated FOA audio, we conduct a subjective listening study with 14 participants. We construct an evaluation set of 100 FOA clips from real recordings and realistic acoustic environments~\citep{shimada2023starss23, heittola_2018_1228142, sharath_adavanne_2019_2636586}, and manually annotate their acoustic events and spatial trajectories as generation conditions. Participants rate each sample from 1 (worst) to 5 (best) along three dimensions: acoustic-event consistency (MOS-Event), spatial-trajectory consistency (MOS-Trajectory), and audio realism (MOS-Realism). As shown in Table~\ref{tab:subjective_eval}, real FOA recordings provide a perceptual upper-bound reference, with MOS scores of 4.89, 4.56, and 4.79, respectively. Among the generation methods, PhysWave achieves substantially higher subjective scores than the compared models across all three dimensions.

\paragraph{LLM parser evaluation.}
We evaluate the LLM spatial parser that converts natural-language spatial descriptions into waypoint-caption conditions. As shown in Table~\ref{tab:parser_eval}, using GPT-4o to score spatial faithfulness and semantic preservation, we find that Qwen3.5-4B performs best among the evaluated open-source parsers, with $3.88 \pm 0.28$ and $4.72 \pm 0.17$, respectively. We therefore use it for descriptive control, while the same waypoint-caption representation also supports direct waypoint input. The detailed protocol is provided in Appendix~\ref{app:parser_eval}.

\begin{figure}[t]
    \centering
    \includegraphics[width=0.48\textwidth]{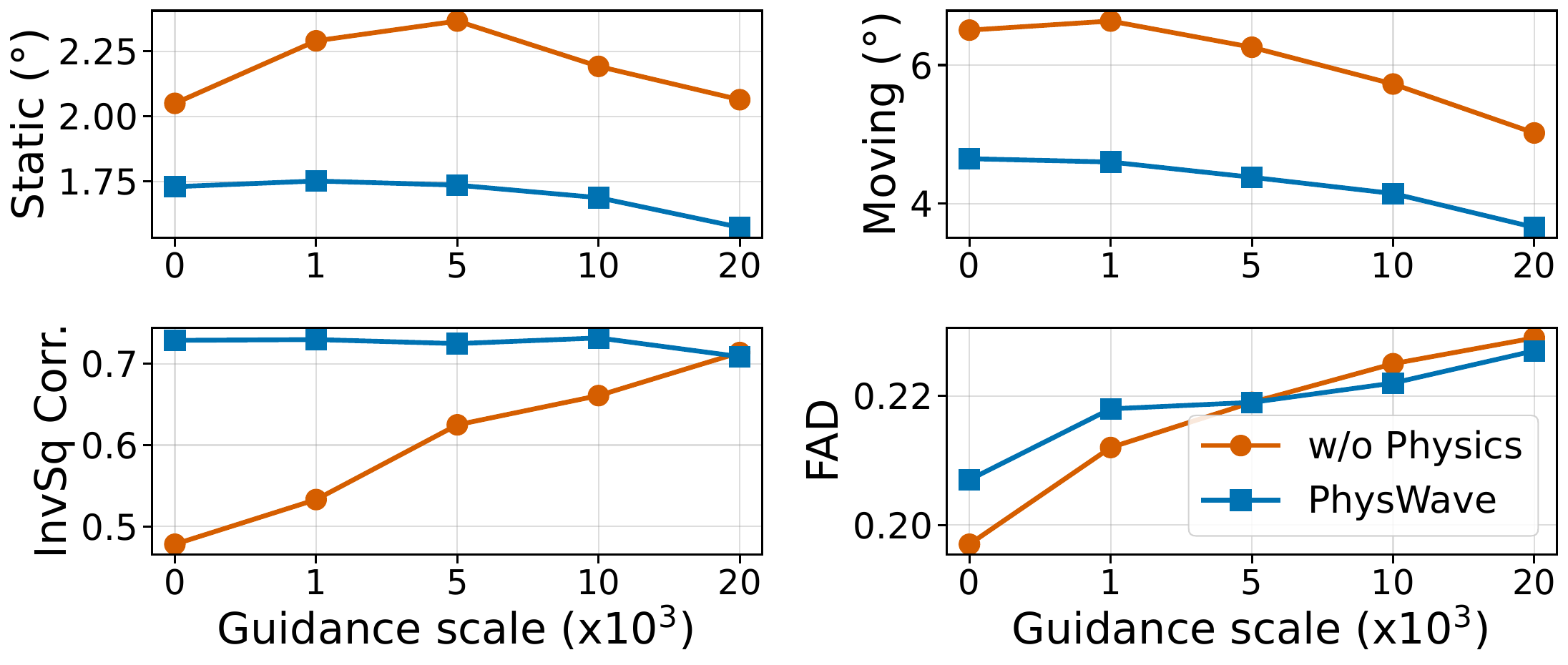}
    \caption{Effect of inference-time physics guidance.}
    \label{fig:guidance_sampling}
\end{figure}

\paragraph{Inference-time physics guidance.}
We further examine whether the same physics-consistency losses can guide sampling without retraining. At selected denoising steps, we compute the differentiable losses ($\mathcal{L}_{\mathrm{dist}}$ and $\mathcal{L}_{\mathrm{dir}}$), and use their gradients to steer the sample toward spatially consistent FOA audio. As shown in Figure~\ref{fig:guidance_sampling}, larger guidance scales generally reduce static and moving angular errors and improve InvSq Corr., especially for ``w/o physics''. FAD$_{\mathrm{CLAP}}$ increases mildly at larger scales, indicating a trade-off between spatial correction and audio quality. These results show that the proposed physics-consistency losses can also serve as inference-time guidance for spatial refinement.

\paragraph{Visualization.}
Figure~\ref{fig:qualitative_trajectory} visualizes source trajectories estimated from generated FOA audio, comparing PhysWave with and without physics-consistency losses during training. Without physics losses, the generated trajectories roughly follow the target motion but show noticeable deviations, especially for static localization and circular motion. In contrast, PhysWave produces trajectories that better align with the ground truth across all motion types. This qualitative trend is consistent with Table~\ref{tab:main_ablation}, where the physics losses reduce angular error and improve distance consistency. Additional trajectory visualizations are provided in Appendix~\ref{app:visualization}.

\section{Conclusion}
We presented \textit{PhysWave}, a physics-guided latent diffusion model for controllable text-to-FOA generation. PhysWave combines waypoint-caption conditioning with differentiable direction and distance priors, improving spatial consistency while maintaining competitive audio quality. 
% The same priors also support training-free inference-time refinement, showing the value of explicit acoustic priors for spatial audio generation.
The same priors also support training-free inference-time refinement, further demonstrating the benefits of explicit acoustic priors for spatial audio generation.

\section*{Limitations}
PhysWave currently focuses on single-source trajectory control in a free-field setting. It does not model room effects such as reverberation, occlusion, and multi-path propagation, which require additional acoustic modeling beyond the direct-path simulation used in this work. Extending physics-guided FOA generation to multi-source scenes and realistic room acoustics is a useful direction for future work.

\section*{Acknowledgments}
The work of Y. Ling, C. Huang, X. Yang and M. Pan were supported in part by the US National Science Foundation under grants CNS-2107057, CNS-2318664, CSR-2403249, and CNS-2431596. This work used Jetstream2 at Indiana University through allocation CIS261116 from the Advanced Cyberinfrastructure Coordination Ecosystem: Services \& Support (ACCESS) program~\citep{boerner2023access}, which is supported by U.S. National Science Foundation grants \#2138259, \#2138286, \#2138307, \#2137603, and \#2138296.

\bibliography{custom}

\begin{table*}[t]
\centering
\small
\setlength{\tabcolsep}{4.2pt}
\renewcommand{\arraystretch}{1.18}
\resizebox{\textwidth}{!}{%
\begin{tabular}{@{}lcccccc@{}}
\toprule
\multirow{2}{*}{\textbf{Method}}
& \multirow{2}{*}{\textbf{Format}}
& \multirow{2}{*}{\textbf{Conditioning}}
& \multicolumn{2}{c}{\textbf{Spatial Source}}
& \multirow{2}{*}{\textbf{Trajectory Control}}
& \multirow{2}{*}{\textbf{Physics-Guided}} \\
\cmidrule(lr){4-5}
& & & \textbf{Static} & \textbf{Moving} & & \\
\midrule
AudioLDM~\citep{liu2023audioldm}
& Mono & Text & -- & -- & -- & \ding{55} \\
AudioLDM 2~\citep{liu2024audioldm}
& Mono & Text & -- & -- & -- & \ding{55} \\
Make-An-Audio~\citep{huang2023make}
& Mono & Text & -- & -- & -- & \ding{55} \\
Make-An-Audio 2~\citep{huang2023make2}
& Mono & Text & -- & -- & -- & \ding{55} \\
TANGO~\citep{ghosal2023tango}
& Mono & Text & -- & -- & -- & \ding{55} \\
TANGO 2~\citep{majumder2024tango}
& Mono & Text & -- & -- & -- & \ding{55} \\
\midrule
Stable Audio Open~\citep{evans2025stable}
& Stereo & Text & -- & -- & -- & \ding{55} \\
SpatialSonic~\citep{sun2024both}
& Stereo & Text/Image + Azimuth & \ding{51} & \ding{51} & Azimuth trajectory & \ding{55} \\
AudioSpa~\citep{feng2025audiospa}
& Binaural & Text + Mono Ref. & \ding{51} & \ding{55} & -- & \ding{55} \\
DualSpec~\citep{zhao2026dualspec}
& Binaural & Text & \ding{51} & \ding{55} & -- & \ding{55} \\
TAS~\citep{pan2025wild}
& Binaural & Text + Mono Ref. & \ding{51} & \ding{51} & Flexible location & \ding{55} \\
\midrule
ViSAGe~\citep{kim2025visage}
& FOA & Video & \ding{51} & \ding{51} & Video-driven & \ding{55} \\
OmniAudio~\citep{liu2025omniaudiogeneratingspatialaudio}
& FOA & 360 Video & \ding{51} & \ding{51} & Video-driven & \ding{55} \\
Diff-SAGe~\citep{kushwaha2025diff}
& FOA & Category + Location & \ding{51} & \ding{55} & -- & \ding{55} \\
ImmerseDiffusion~\citep{heydari2025immersediffusion}
& FOA & Text / Spatial Params. & \ding{51} & \ding{55} & -- & \ding{55} \\
SonicMotion~\citep{templin2025generating}
& FOA & Text / Spatial Params. & \ding{51} & \ding{51} & Circular & \ding{55} \\
\midrule
\rowcolor{blue!10}
\textbf{PhysWave (Ours)}
& \textbf{FOA}
& \textbf{Text + Waypoints}
& \textbf{\ding{51}}
& \textbf{\ding{51}}
& \textbf{General waypoints}
& \textbf{\ding{51}} \\
\bottomrule
\end{tabular}
}
\caption{Functional comparison with prior audio generation methods. Mono, Stereo, Binaural, and FOA denote output formats. \ding{51} means supported, \ding{55} means not supported, and ``--'' means not applicable. For SpatialSonic, trajectory control refers to azimuth-level control, while PhysWave supports listener-relative waypoint trajectories with azimuth, elevation, and distance. Physics-guided means that explicit acoustic priors are used during generation.}
\label{tab:functional_comparison}
\end{table*}

\begin{table*}[t]
\centering
\small
\setlength{\tabcolsep}{5pt}
\renewcommand{\arraystretch}{1.12}
\resizebox{\textwidth}{!}{%
\begin{tabular}{lcll}
\toprule
\textbf{Family} & \textbf{Ratio} & \textbf{Sampling Range} & \textbf{Description} \\
\midrule
Static
& $50.0\%$
& $\theta\in[-180^\circ,180^\circ]$, $\phi\in[-35^\circ,35^\circ]$,
$r\in[0.5,5]$ m
& Fixed direction and distance \\
Linear pass-by
& $\approx16.7\%$
& $v\in[1,25]$ m/s, $d_{\min}\in[1,8]$ m
& Coupled direction and distance changes \\
Circular
& $\approx16.7\%$
& $r\in[0.5,5]$ m, angular span $\ge 30^\circ$
& Direction changes at near-constant distance \\
Approach/recede
& $\approx16.7\%$
& $v\in[2,6]$ m/s, $r\in[1,60]$ m
& Distance changes with fixed direction \\
\bottomrule
\end{tabular}
}
\caption{Trajectory sampling configuration. Each retained monaural clip is rendered once as a static source and once as a moving source.}
\label{tab:trajectory_sampling}
\end{table*}

\appendix
\section{Comparison with Prior Works}
\label{app:comparison}
Table~\ref{tab:functional_comparison} compares PhysWave with prior audio generation methods in terms of output format, conditioning input, spatial source support, trajectory control, and physics guidance. Compared with monaural and stereo/binaural methods, PhysWave directly generates FOA spatial audio. Compared with existing FOA generators, PhysWave supports waypoint-based trajectory control for both static and moving sources, going beyond the static-source setting of ImmerseDiffusion and the circular-motion setting of SonicMotion. PhysWave also integrates explicit acoustic priors into the generation process to improve spatial consistency.

\section{Dataset Construction Details}
\label{app:dataset}
\subsection{Source Audio Preparation}
We construct the source pool from three public captioned audio datasets: AudioCaps~\citep{audiocaps}, WavCaps~\citep{mei2023wavcaps}, and Clotho~\citep{drossos2020clotho}. Since public captioned audio can contain noisy captions, weak audio-text matches, or overlapping events, we first prioritize clips whose captions describe a dominant acoustic event, following the single-source pre-selection strategy used in prior spatial-audio dataset construction~\citep{sun2024both}. All retained clips are converted to mono and resampled to $16$ kHz. We then apply energy-based activity detection to extract a $10$-second segment with sufficient active content. Clips with less than $1$ second of active content are discarded. If a clip is shorter than $10$ seconds, we pad it with zeros. If it is longer than $10$ seconds, we select the $10$-second window containing the largest number of active frames.
We apply EBU R128 loudness normalization to a target level of $-14$ LUFS and clip the normalized waveform to $[-1,1]$. We then compute audio-caption similarity using a CLAP model and discard clips with scores below $0.3$. For Clotho, where each clip has up to five captions, we score all captions and keep the highest-scoring one. As shown in Table~\ref{tab:data_processing_stats}, the final source pool contains $157{,}409$ training clips and $5{,}260$ test clips. Each source clip is rendered once as a static source and once as a moving source, resulting in $314{,}818$ training FOA examples and $10{,}520$ test FOA examples.

\begin{table}[t]
\centering
\small
\setlength{\tabcolsep}{2.5pt}
\renewcommand{\arraystretch}{1.25}
\resizebox{\columnwidth}{!}{%
\begin{tabular}{lrrrr}
\toprule
\multirow{2}{*}{\textbf{Data Source}}
& \multicolumn{2}{c}{\textbf{Train}}
& \multicolumn{2}{c}{\textbf{Test}} \\
\cmidrule(lr){2-3}\cmidrule(lr){4-5}
& \textbf{Clips} & \textbf{Duration (h)}
& \textbf{Clips} & \textbf{Duration (h)} \\
\midrule
AudioCaps & 46,696  & 129.7 & 4,239 & 11.8 \\
WavCaps   & 107,850 & 299.6 & --    & --   \\
Clotho    & 2,863   & 8.0   & 1,021 & 2.8  \\
\midrule
Total monaural  & 157,409 & 437.2 & 5,260  & 14.6 \\
Rendered FOA & 314,818 & 874.5 & 10,520 & 29.2 \\
\bottomrule
\end{tabular}%
}
\caption{Statistics of source data after preprocessing. Each retained monaural clip is rendered twice, once as a static source and once as a moving source.}
\label{tab:data_processing_stats}
\end{table}

\begin{figure*}[t]
\centering
\begin{tcolorbox}[
    enhanced,
    width=0.82\textwidth,
    colback=white,
    colframe=gray!65,
    boxrule=0.5pt,
    sharp corners,
    arc=0pt,
    left=8pt,
    right=8pt,
    top=6pt,
    bottom=6pt,
    borderline={0.6pt}{0pt}{gray!65,dotted}
]
\begin{lstlisting}[style=promptbox, escapechar=|]
{
  |\textcolor{red!75!black}{\textbf{"Task"}}|:
  "You are a spatial audio caption parser. Convert the input scene text into strict JSON for spatial audio generation. Do not output waypoint arrays; a deterministic program will convert the parsed trajectory into 10 waypoints.",

  |\textcolor{red!75!black}{\textbf{"Coordinate system"}}|:
  "az in [-180,180], 0=front, +90=left, -90=right, +/-180=back; el is elevation in degrees, 0=horizon; r is distance in meters; time is in seconds.",

  |\textcolor{red!75!black}{\textbf{"Output schema"}}|: {
    "duration": 10.0,
    "events": [{
      "text": "<acoustic content only, no spatial words>",
      "t_start": <float>,
      "t_end": <float>,
      "trajectory": {
        "type": "static | linear | arc | approach | recede",
        "start": {"az": <float>, "el": <float>, "r": <float>},
        "end": {"az": <float>, "el": <float>, "r": <float>},
        "control_points": [
          {"time": <float>, "az": <float>, "el": <float>, "r": <float>}
        ],
        "direction": "clockwise | counterclockwise | null",
        "turns": <float>,
        "speed": "slow | medium | fast"
      },
      "inferred": [<string>]
    }]
  },

  |\textcolor{red!75!black}{\textbf{"Rules"}}|: [
    "Output JSON only.",
    "Use static for fixed sources; linear for motion from one position to another.",
    "Use approach/recede for distance change; use arc for circular or orbiting motion.",
    "If time is unspecified, use t_start=0.0 and t_end=10.0.",
    "Defaults: az=0, el=0, normal r=8.0, close r=2.0, very close r=1.0, far r=25.0.",
    "If a value is inferred, list its field path in inferred."
  ],

  |\textcolor{red!75!black}{\textbf{"Input"}}|: "{SCENE_TEXT}"
}
\end{lstlisting}
\end{tcolorbox}
\vspace{-0.5em}
\caption{Prompt used by the LLM spatial parser. The parser outputs a compact trajectory representation, which is converted into the $M=10$ waypoint condition used by the diffusion model.}
\label{fig:spatial_parser_prompt}
\end{figure*}

\subsection{Trajectory Sampling}
For each retained monaural clip, we render two spatial versions: one static source and one moving source. The static version keeps a fixed azimuth, elevation, and distance throughout the clip. The moving version is sampled from three motion families: linear pass-by, circular motion, and approach/recede motion. All interpolated trajectories are stored as frame-level trajectory samples $(t,\theta,\phi,r)$ at $0.1$ s intervals, where $\theta$ is azimuth, $\phi$ is elevation, and $r$ is the source-listener distance. Table~\ref{tab:trajectory_sampling} summarizes the sampling configuration. Since each clip produces one static and one moving example, static sources account for $50\%$ of the rendered FOA samples. The moving examples are split nearly uniformly across the three motion families, so each moving family contributes about $16.7\%$ of all rendered FOA samples.

\section{LLM Spatial Parser}
\subsection{Parser Prompt}
We use an LLM spatial parser to convert a natural-language spatial caption into a structured intermediate representation. The parser separates the non-spatial acoustic content from spatial trajectory attributes and outputs them in JSON format. A deterministic post-processing module then converts the parsed trajectory into the $M=10$ waypoint condition used by the diffusion model. The full prompt is shown in Figure~\ref{fig:spatial_parser_prompt}.

\subsection{Example Parser Outputs}
\label{app:parser_examples}

Figure~\ref{fig:parser_examples} shows example parser outputs. For compactness, we only show the main parsed fields.

\begin{figure*}[t]
\centering
\begin{minipage}{0.48\textwidth}
\begin{tcolorbox}[
    enhanced,
    colback=white,
    colframe=gray!65,
    boxrule=0.5pt,
    sharp corners,
    arc=0pt,
    left=6pt,
    right=6pt,
    top=5pt,
    bottom=5pt,
    borderline={0.6pt}{0pt}{gray!65,dotted}
]
\begin{lstlisting}[style=promptbox, escapechar=|]
{
  |\textcolor{red!75!black}{\textbf{"Input"}}|:
  "Ocean waves crashing as water trickles and splashes,
   approaching from the left, moving from farther away
   to a closer distance.",

  |\textcolor{red!75!black}{\textbf{"Output"}}|: {
    "text": "Ocean waves crashing as water trickles and splashes.",
    "trajectory": {
      "type": "approach",
      "start": {"az": 90, "el": 0, "r": 25},
      "end": {"az": 90, "el": 0, "r": 2}
    }
  }
}
\end{lstlisting}
\end{tcolorbox}
\end{minipage}
\hfill
\begin{minipage}{0.48\textwidth}
\begin{tcolorbox}[
    enhanced,
    colback=white,
    colframe=gray!65,
    boxrule=0.5pt,
    sharp corners,
    arc=0pt,
    left=6pt,
    right=6pt,
    top=5pt,
    bottom=5pt,
    borderline={0.6pt}{0pt}{gray!65,dotted}
]
\begin{lstlisting}[style=promptbox, escapechar=|]
{
  |\textcolor{red!75!black}{\textbf{"Input"}}|:
  "Burping and a man speaking, passing from the
   front-right to the back-left, passing at a normal
   distance.",

  |\textcolor{red!75!black}{\textbf{"Output"}}|: {
    "text": "Burping and a man speaking.",
    "trajectory": {
      "type": "linear",
      "start": {"az": -45, "el": 0, "r": 8},
      "end": {"az": 135, "el": 0, "r": 8}
    }
  }
}
\end{lstlisting}
\end{tcolorbox}
\end{minipage}
\vspace{-0.5em}
\caption{Example outputs of the LLM spatial parser. Each output keeps the
non-spatial acoustic content and converts the spatial description into a
parametric trajectory.}
\label{fig:parser_examples}
\end{figure*}

\section{LLM-as-a-Judge Evaluation}
\label{app:parser_eval}
We use an LLM-as-a-judge protocol to evaluate the spatial parser. For each spatial caption, a candidate parser generates a structured trajectory representation, which is converted into a fixed-length waypoint condition by deterministic post-processing. The judge receives the original spatial caption, the parsed acoustic caption, and the converted waypoints. It then assigns two integer scores from 1 to 5: \emph{spatial faithfulness}, which measures whether the waypoints match the described spatial layout and motion, and \emph{semantic preservation}, which measures whether the parsed caption preserves the acoustic content without spatial leakage. We evaluate 200 randomly sampled spatial captions using GPT-4o with temperature set to 0. All candidate parsers are evaluated on the same caption set using the same judge prompt, and each example is scored once. The judge is instructed to output only the two integer scores and a short justification, where the justification is used only for inspection and is not included in the reported metrics. The full judge prompt is shown in Figure~\ref{fig:llm_judge_prompt}.

\begin{figure}[t]
\centering
\begin{tcolorbox}[
    enhanced,
    width=\columnwidth,
    colback=white,
    colframe=gray!65,
    boxrule=0.5pt,
    sharp corners,
    arc=0pt,
    left=6pt,
    right=6pt,
    top=5pt,
    bottom=5pt,
    borderline={0.6pt}{0pt}{gray!65,dotted}
]
\begin{lstlisting}[style=promptbox, escapechar=|, basicstyle=\ttfamily\tiny]
{
  |\textcolor{red!75!black}{\textbf{"Task"}}|:
  "You are evaluating an LLM parser for a spatial audio system.",

  |\textcolor{red!75!black}{\textbf{"Original spatial caption"}}|:
  "{SPATIAL_CAPTION}",

  |\textcolor{red!75!black}{\textbf{"Parser output"}}|:
  "- Text caption: {PARSED_CAPTION}
   - Waypoints (t in seconds, azimuth/elevation in degrees,
     distance in meters): {WAYPOINTS}",

  |\textcolor{red!75!black}{\textbf{"Coordinate convention"}}|:
  "- azimuth 0 deg = front
   - +45 deg = front-left; +90 deg = left; +135 deg = back-left
   - -45 deg = front-right; -90 deg = right; -135 deg = back-right
   - +/-180 deg = behind/back
   - elevation + is above and elevation - is below
   - distance defaults: very close ~= 1m, close ~= 2m,
     normal ~= 8m, far ~= 25m

   Use this convention exactly; do not swap left and right.
   If a caption does not specify exact timing, do not penalize
   the parser for using the full 0-10s range.",

  |\textcolor{red!75!black}{\textbf{"Scoring"}}|:
  "Score each criterion on a 1-5 integer scale.",

  |\textcolor{red!75!black}{\textbf{"Spatial faithfulness"}}|:
  "5 = all spatial and temporal details are accurate;
   4 = minor inaccuracies but overall faithful;
   3 = one major spatial aspect is wrong;
   2 = multiple spatial aspects are wrong;
   1 = waypoints do not match the described layout.",

  |\textcolor{red!75!black}{\textbf{"Semantic preservation"}}|:
  "5 = acoustic content fully preserved with no spatial leakage;
   4 = acoustic content preserved with minor spatial leakage;
   3 = some acoustic content is lost or spatial leakage is clear;
   2 = acoustic content is strongly degraded or spatial leakage is severe;
   1 = acoustic content is lost or mixed with spatial details.",

  |\textcolor{red!75!black}{\textbf{"Output format"}}|:
  "Return only this JSON object:
   {
     \"spatial_faithfulness\": <int 1-5>,
     \"semantic_preservation\": <int 1-5>,
     \"justification\": \"<one short sentence per criterion>\"
   }"
}
\end{lstlisting}
\end{tcolorbox}
\vspace{-0.5em}
\caption{Prompt used for LLM-as-a-judge parser evaluation. The judge compares the original spatial caption with the parsed acoustic caption and converted waypoints.}
\label{fig:llm_judge_prompt}
\end{figure}

\section{Model Details}
\label{app:model}

\subsection{FOA VAE Training}
\label{app:foa_vae}
We train a VAE-based neural audio codec to compress FOA waveforms into a continuous latent space for diffusion modeling. The autoencoder follows a DAC-style 1D convolutional architecture, with the input and output channels changed to the four FOA components $(W,X,Y,Z)$. All audio is represented at $16$ kHz. We train the VAE on $2.05$-second crops, with a downsampling ratio of $1024$ and a latent channel dimension of $64$. After training, the VAE is frozen and used as the latent encoder and decoder for the diffusion model. Related work has also studied FOA-specific neural audio representations for spatially consistent tokenization~\citep{sudarsanam2025foa}.

\paragraph{Training objective.}
The VAE objective combines spectral reconstruction, adversarial training, latent regularization, and FOA spatial consistency:
\begin{equation}
\begin{aligned}
\mathcal{L}_{\mathrm{VAE}}
=&\;
\frac{\lambda_{\mathrm{mrstft}}}{4}
\sum_{c\in\{W,X,Y,Z\}}
\mathcal{L}_{\mathrm{mrstft}}^{c}
+
\lambda_{\mathrm{kl}}\mathcal{L}_{\mathrm{kl}}
\\
&+
\lambda_{\mathrm{adv}}\mathcal{L}_{\mathrm{adv}}
+
\lambda_{\mathrm{fm}}\mathcal{L}_{\mathrm{fm}}
+
\lambda_{\mathrm{dir}}\mathcal{L}_{\mathrm{dir}} .
\end{aligned}
\label{eq:vae_loss}
\end{equation}
Here, $\mathcal{L}_{\mathrm{mrstft}}^{c}$ denotes the multi-resolution STFT loss computed on FOA channel $c\in\{W,X,Y,Z\}$. The terms $\mathcal{L}_{\mathrm{adv}}$ and $\mathcal{L}_{\mathrm{fm}}$ are the adversarial and feature-matching losses from the discriminator, and $\mathcal{L}_{\mathrm{kl}}$ regularizes the VAE bottleneck. To better preserve FOA spatial structure, we add $\mathcal{L}_{\mathrm{dir}}$ to improve spatial reconstruction consistency. We set $\lambda_{\mathrm{mrstft}}=1.0$, $\lambda_{\mathrm{adv}}=0.1$, $\lambda_{\mathrm{fm}}=5.0$, $\lambda_{\mathrm{kl}}=10^{-6}$, and $\lambda_{\mathrm{dir}}=0.1$.

\paragraph{Optimization.}
We train the autoencoder on $2.05$-second crops from the training split. The autoencoder and discriminator are optimized with AdamW using betas $(0.8,0.99)$ and weight decay $10^{-3}$. The autoencoder uses learning rate $1\times 10^{-4}$, while the discriminator uses learning rate $2\times 10^{-4}$. Both learning rates are decayed with an exponential scheduler with decay factor $0.999996$. The autoencoder is trained on two NVIDIA RTX 3090 GPUs with a batch size of 12 for $120$K steps.

\paragraph{Reconstruction quality.}
Table~\ref{tab:vae_recon} reports FOA VAE reconstruction quality on the test split. The VAE preserves waveform quality while maintaining a low angular reconstruction error of $1.92^\circ$.

\begin{table}[t]
\centering
\small
\setlength{\tabcolsep}{5pt}
\caption{FOA VAE reconstruction quality.}
\label{tab:vae_recon}
\resizebox{\columnwidth}{!}{%
\begin{tabular}{lccccc}
\toprule
\textbf{Model}
& \textbf{STFT}$\downarrow$
& \textbf{Mel}$\downarrow$
& \textbf{L1}$(\theta)\downarrow$
& \textbf{L1}$(\phi)\downarrow$
& $\Delta_{\textbf{angle}}\downarrow$ \\
\midrule
FOA VAE
& 1.44
& 1.11
& 0.87$^\circ$
& 1.45$^\circ$
& 1.92$^\circ$ \\
\bottomrule
\end{tabular}%
}
\end{table}

\subsection{Diffusion Training}
We train the latent diffusion model on top of the frozen FOA VAE. Each $10$-second FOA clip is encoded into a latent sequence with $64$ channels. The denoising network is a DiT with hidden dimension $768$, $24$ transformer layers, and $12$ attention heads, and predicts the velocity target. Text captions are encoded by T5-base with a maximum length of $128$ tokens. The interpolated trajectory features are encoded by a Q-Former-style encoder with 160 input frames, 16 learnable query tokens, hidden size 768, 8 attention heads, and MLP hidden size 256. The text tokens, waypoint, and temporal tokens are used as conditioning inputs to the DiT.

\paragraph{Optimization.}
We train the diffusion model with AdamW using a learning rate $1\times 10^{-5}$, betas $(0.9,0.99)$, and weight decay $10^{-3}$. The learning rate follows a linear warm-up followed by cosine decay to $1\times 10^{-6}$. EMA weights with decay $0.9999$ are maintained during training. Classifier-free conditioning dropout is applied with probability $0.1$. The model is trained on one NVIDIA H100 GPU, with a batch size of $64$ and bfloat16 mixed precision. The reported PhysWave checkpoint is trained for $128$K optimization steps.

\paragraph{Physics-consistency loss.}
In addition to the latent denoising loss, we apply a physics-consistency loss to decoded waveform estimates during training:
\begin{equation}
\mathcal{L}_{\mathrm{phys}}
=
\lambda_{\mathrm{dir}}\mathcal{L}_{\mathrm{dir}}
+
\lambda_{\mathrm{dist}}\mathcal{L}_{\mathrm{dist}} .
\end{equation}
We use $\lambda_{\mathrm{dir}}=1.0$ and $\lambda_{\mathrm{dist}}=0.05$. The loss is computed on $8$ sampled denoising examples per batch, using $40$ ms analysis frames. The loss is applied from the start of training and weighted by the denoising SNR.

\section{Evaluation Metric Definitions}
\label{app:metric}
We provide implementation details for the evaluation metrics used in Section~\ref{sec:experiments}. All audio quality metrics are computed on the omnidirectional $W$ channel of the generated FOA signal, since standard text-to-audio metrics are defined for single-channel or general audio evaluation.

\paragraph{Audio quality metrics.}
We follow standard text-to-audio evaluation protocols and report Fréchet Distance (FD), CLAP-based Fréchet Audio Distance (FAD$_{\mathrm{CLAP}}$), Inception Score (IS), Kullback--Leibler divergence (KL), and CLAP score. FD, IS, and KL are computed using a pretrained PANNs classifier~\citep{kong2020panns}. Specifically, FD measures the Fréchet distance between generated and reference audio embeddings, while IS and KL are computed from the class-posterior distributions predicted by PANNs. FAD$_{\mathrm{CLAP}}$ is computed in the CLAP embedding space~\citep{wu2023large}. CLAP score is computed as the cosine similarity between the CLAP audio embedding of the generated $W$ channel and the CLAP text embedding of the input caption.

\paragraph{Spatial fidelity metrics.}
We evaluate spatial fidelity using direction consistency and distance consistency. For direction consistency, we estimate the frame-level direction of arrival (DoA) from the generated FOA signal. For each frame $k$, we compute the FOA intensity vector
\begin{equation}
\hat{\mathbf{I}}_k =
\left[
\langle W_k X_k\rangle,\;
\langle W_k Y_k\rangle,\;
\langle W_k Z_k\rangle
\right]^\top ,
\end{equation}
where $\langle\cdot\rangle$ denotes averaging within the frame. The estimated azimuth and elevation are
\begin{equation}
\begin{aligned}
\hat{\theta}_k
&=
\operatorname{atan2}(\hat{I}_{y,k},\hat{I}_{x,k}),\\
\hat{\phi}_k
&=
\operatorname{atan2}
\left(
\hat{I}_{z,k},
\sqrt{\hat{I}_{x,k}^2+\hat{I}_{y,k}^2}
\right).
\end{aligned}
\end{equation}
Given the ground-truth direction $(\theta_k,\phi_k)$, we compute the spherical angular error using the haversine formula. Let
\begin{equation}
a_k
=
\sin^2\left(\frac{\Delta\phi_k}{2}\right)
+
\cos(\phi_k)\cos(\hat{\phi}_k)
\sin^2\left(\frac{\Delta\theta_k}{2}\right),
\end{equation}
where $\Delta\theta_k=\hat{\theta}_k-\theta_k$ and
$\Delta\phi_k=\hat{\phi}_k-\phi_k$. The DoA error is
\begin{equation}
\Delta_{\mathrm{angle},k}
=
2\cdot
\operatorname{atan2}
\left(
\sqrt{a_k},\sqrt{1-a_k}
\right).
\end{equation}
We report the mean $\Delta_{\mathrm{angle}}$ in degrees over active frames.

For distance consistency, we evaluate whether the generated signal follows the trajectory-conditioned inverse-square attenuation profile defined by the ground-truth distance trajectory. We first compute the generated $W$-channel energy at each frame:
\begin{equation}
E_k = \langle \hat{W}_k^2\rangle .
\end{equation}

\begin{figure*}[t]
    \centering
    \includegraphics[width=0.85\textwidth]{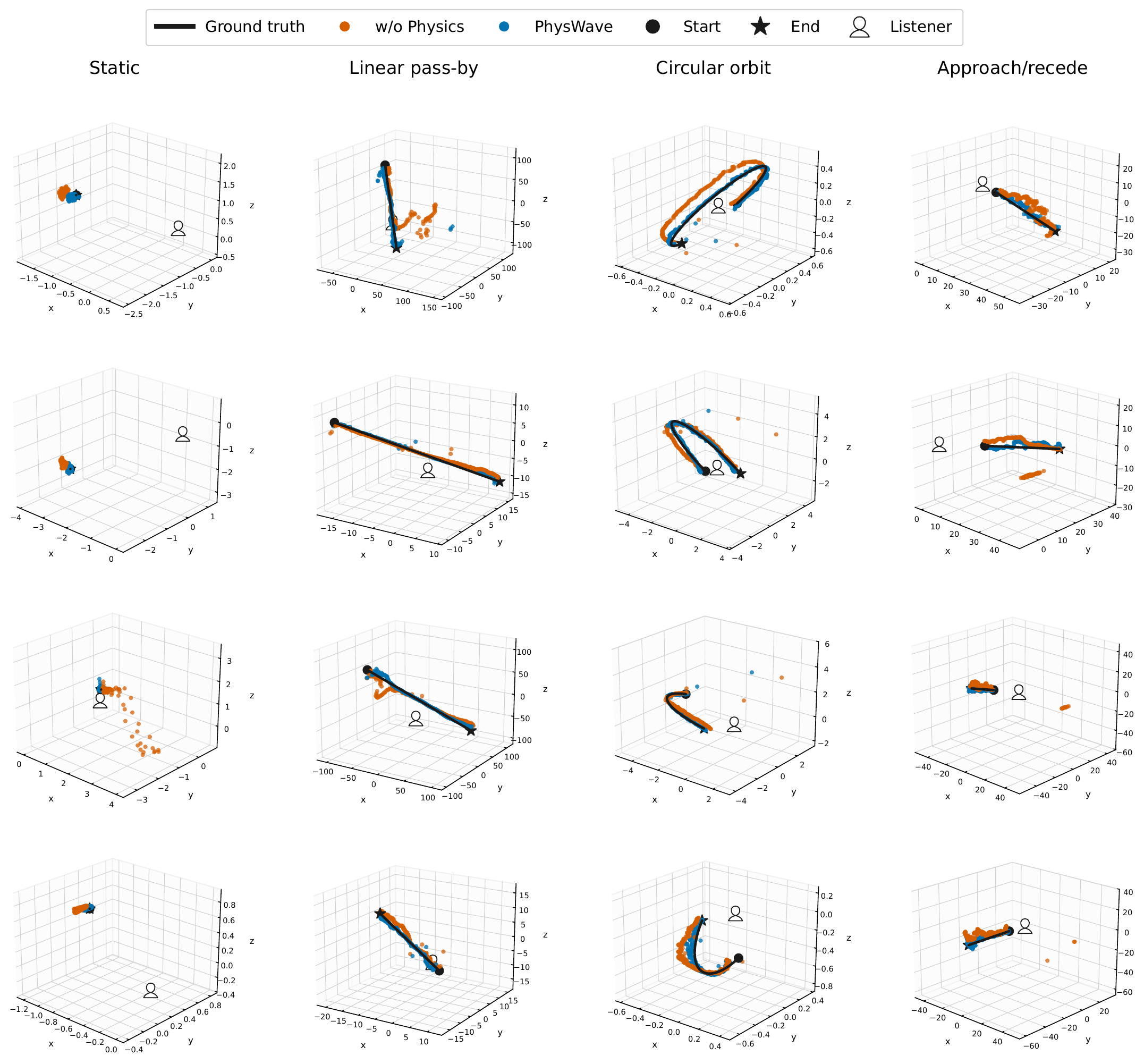}
    \caption{More qualitative trajectory comparison between PhysWave with and without physics-consistency losses.}
    \label{fig:more_trajectory}
\end{figure*}

We define active frames as
\begin{equation}
\mathcal{K}=\{k \mid E_k \ge 10^{-3}\max_j E_j\},
\end{equation}
and the target inverse-square profile as
\begin{equation}
q_k = \frac{1}{r_k^2},
\end{equation}
where $r_k$ is the ground-truth source-listener distance. To remove source-dependent absolute loudness, we normalize log-energy as
\begin{equation}
\widetilde{u}_k =
\log(u_k+\epsilon)
-
\frac{1}{|\mathcal{K}|}
\sum_{j\in\mathcal{K}}\log(u_j+\epsilon),
\end{equation}
where normalization is computed over active frames. We define InvSqErr as
\begin{equation}
\mathrm{InvSqErr}
=
\frac{10}{\log 10}
\sqrt{
\frac{1}{|\mathcal{K}|}
\sum_{k\in\mathcal{K}}
\left(
\widetilde{E}_k-\widetilde{q}_k
\right)^2
},
\end{equation}
and InvSqCorr as
\begin{equation}
\mathrm{InvSqCorr}
=
\operatorname{corr}_{k\in\mathcal{K}}(\widetilde{E}_k,\widetilde{q}_k).
\end{equation}
Lower InvSqErr and higher InvSqCorr indicate better agreement with the specified distance attenuation cue.

\section{Additional Visualizations}
\label{app:visualization}
\begin{figure}[h]
    \centering
    \includegraphics[width=0.48\textwidth]{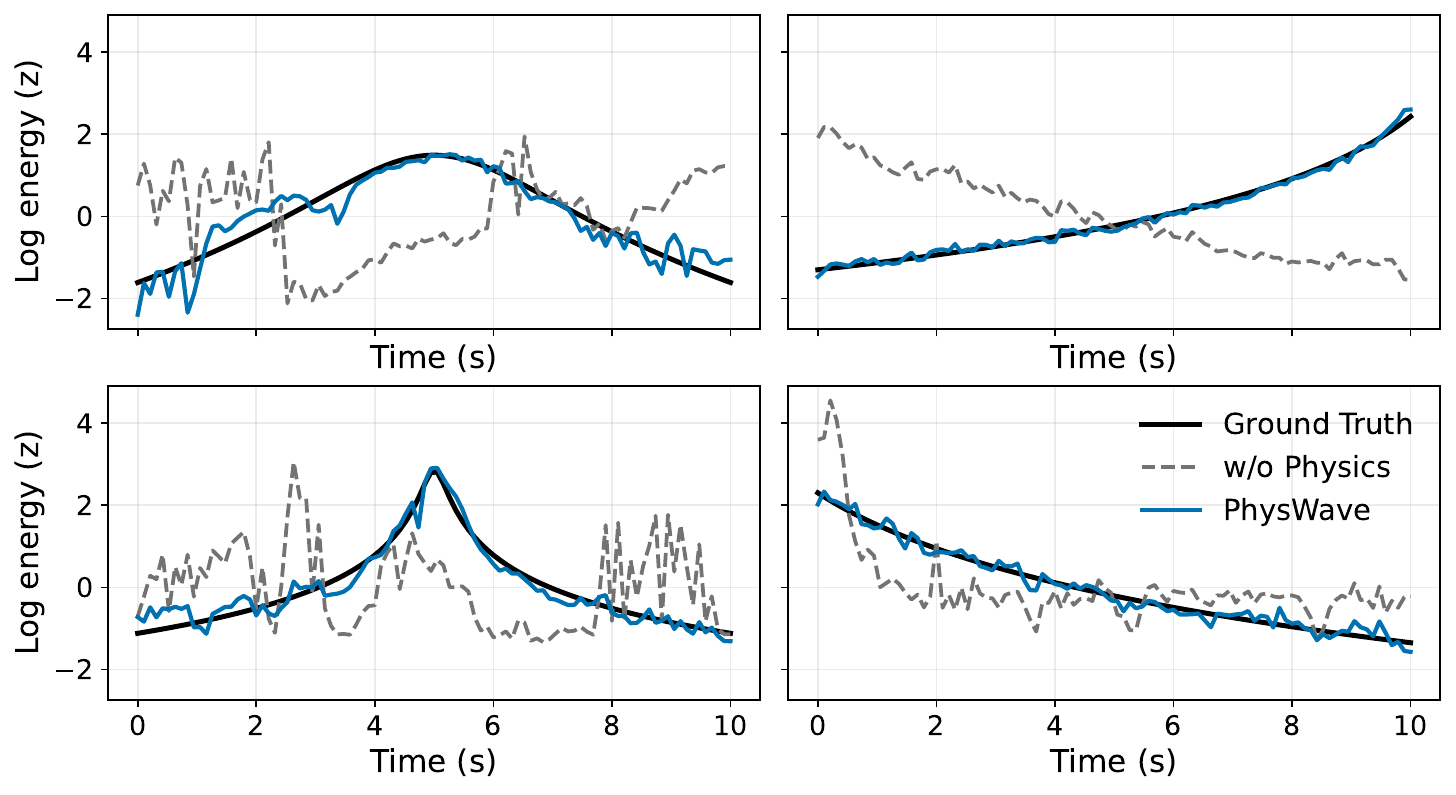}
    \caption{Energy envelope comparison. PhysWave better follows the target inverse-square log-energy profile, while the model without physics losses shows larger temporal deviations.}
    \label{fig:energy_envelope}
\end{figure}

We provide additional visualizations of PhysWave's spatial behavior. Figure~\ref{fig:energy_envelope} compares the generated log-energy envelope with the target inverse-square profile over time. Figure~\ref{fig:more_trajectory} shows additional trajectory examples across the four motion families.

\end{document}